\documentclass[journal]{IEEEtran}
\ifCLASSINFOpdf
  \usepackage[pdftex]{graphicx}
  \graphicspath{{../pdf/}{../jpeg/}}
  \DeclareGraphicsExtensions{.pdf,.jpeg,.png}
\else
\fi
\usepackage{amsmath}
\usepackage{array}
\usepackage{booktabs}
\begin{document}
\title{Physics-Informed Neural Networks for 2D Plane Wave Scattering in Arbitrary Dielectric Structures}

\author{Zheng-Yu~Huang,~\IEEEmembership{Member,~IEEE,}
        Yu~Tian, Jing-Wen~Zhang,
        and~Nicolae-Coriolan~Panoiu,~\IEEEmembership{Member,~IEEE}% <-this % stops a space
\thanks{This work was supported in part by National Key Laboratory Foundation of China under Grant JCKYS2023LD5, Aeronautical Science Foundation of China under Grant 20240018052002, the Fund of Nanjing Major Science and Technology Special Projects under Grant 202405029, Natural Science Foundation of Jiangsu Province under Grant BK20252026, and UK's Engineering and Physical Sciences Research Council under Grant UKRI3174.  \textit{(Corresponding author: Zheng-Yu Huang.)}}% <-this % stops a space
\thanks{Zheng-Yu Huang, Yu Tian, Jing-Wen Zhang are with the College of Electronic and Information Engineering, Nanjing University of Aeronautics and Astronautics, Nanjing 210016, China (e-mail: huangzynj@163.com).}% <-this % stops a space
\thanks{Nicolae-Coriolan Panoiu is with the Department of Electronic and Electrical Engineering, University College London, Torrington Place, London WC1E 7JE, United Kingdom (e-mail: n.panoiu@ucl.ac.uk).}
}


\markboth{Journal of \LaTeX\ Class Files,~Vol.~14, No.~8, August~2015}%
{Shell \MakeLowercase{\textit{et al.}}: Bare Demo of IEEEtran.cls for IEEE Journals}



\maketitle

\begin{abstract}
In this paper, we introduce a meshless physics-informed neural network based computational framework for solving two-dimensional electromagnetic wave scattering in inhomogeneous media. The framework embeds frequency-domain Maxwell equations and radiation boundary conditions directly into the neural network loss function, enabling accurate prediction of scattered fields for both transverse magnetic (TM) and transverse electric (TE) polarizations across various dielectric configurations. Application of the method to single-cylinder, concentric multilayer cylindrical shells, three arbitrarily arranged cylinders, and composite irregular structures demonstrates that for the TM polarization, all relative $L^{2}$ errors mostly remain at particularly low levels of $\le0.1$. For the TE polarization, sharp variations of the dielectric properties of scatterers lead to singularities in the governing equations, which result in decreased accuracy of the method. This challenge is overcome by introducing at dielectric boundaries a hyperbolic-tangent smoothing function. This procedure significantly improves the accuracy of the method, with the corresponding results closely matching the predictions of the finite-difference time-domain method. This framework exhibits stable convergence behavior across all of the investigated configurations, thus confirming its robustness and scalability to complex electromagnetic scattering problems.
\end{abstract}


% \fi
\IEEEpeerreviewmaketitle

\section{Introduction}\label{intro}
\IEEEPARstart{N}{umerical} investigations of wave scattering are widely used in many areas of science and engineering, including radar and remote sensing \cite{He2025}, materials science \cite{Ma2025}\cite{Biswas2025}, nuclear physics and astrophysics \cite{Lu2025}, acoustics \cite{Gu2025}, and optics \cite{Sekulic2025}. For example, in optics, electromagnetic wave scattering is a fundamental phenomenon that plays a central role in the interaction between light and micro/nanostructured materials. Solving the electromagnetic wave scattering problem is crucial to the modeling and design of functional nanostructures, such as ensembles of nanoparticles, photonic crystals, and antenna arrays with subwavelength features \cite{Xiong2024}. Numerical methods for solving the electromagnetic wave scattering problem include the finite-difference time-domain (FDTD) method \cite{Gronnemose2025}, the finite element method (FEM) \cite{Song2025}, the boundary element method (BEM) \cite{Wang2024}, and the multiple-scattering method (MSM) \cite{MMIBook2014}\cite{Sekulic2021}. These methods, based on different discretization schemes of the electromagnetic fields and solution representation strategies, have been widely applied to various problems. However, there is still considerable room to advance these numerical methods in terms of scalability, geometric flexibility, and computational cost, especially when analyzing multiscale problems and highly inhomogeneous structures possessing highly irregular boundaries.

One of the most promising research directions aiming to overcome these challenges brings together numerical methods based on neural network optimization and prediction. In this context, methods based on physics-informed neural networks (PINNs), proposed by Raissi et al. \cite{Raissi2019}, have emerged as a particularly powerful and versatile computational framework. These methods use feedforward neural networks (FNNs) to directly approximate continuous physical fields so that explicit mesh discretization of the computational domain is not required \cite{Lawal2022}. By embedding governing equations, boundary conditions, and initial conditions into the loss function and using automatic differentiation to compute residual terms, they can simultaneously satisfy observational data constraints and enforce physical laws during training \cite{Yu2022}. The meshless nature of this approach and strong scalability make PINN-based methods particularly suitable for complex structures and multiphysics problems. For example, PINNs have been applied to modeling high-speed fluid dynamics \cite{Mao2020}, subsurface structure inversion \cite{Moseley2020}, seismic wave propagation \cite{Song2021}, and acoustic field dynamics in complex domains \cite{Wang2023}.

In the analysis of scattering problems, Nair et al. \cite{Nair2024} combined the acoustic wave superposition principle with PINNs to develop a multiple scattering simulation framework describing arbitrary geometric structures and high-frequency conditions. Moreover, Chen et al. \cite{Chen2020}\cite{Chen2022} integrated electromagnetic physics with observational data to achieve parameter inversion and image reconstruction of nanophotonic structures, whereas Saba et al. \cite{Saba2022} trained PINNs as a forward model to rapidly predict scattering fields for various samples, thus providing solutions for optical forward and inverse problems. In addition, Medvedev et al. \cite{Medvedev2025} applied PINNs to three-dimensional diffraction modeling of optical metasurfaces, successfully predicting the scattered electric fields pertaining to different meta-atom materials and geometric structures. In a related work, Zheng et al. \cite{Zheng2025} proposed a hybrid physics-data-driven training method for electromagnetic scattering that combined Maxwell residual constraints with small amounts of labeled field data; as compared to purely physics-informed neural networks, this method converged faster and achieved higher numerical accuracy.

Despite these early successes, currently available PINN-based numerical methods for analysis of electromagnetic scattering problems still require further significant advances. For example, in a high-frequency electromagnetic scattering regime with strongly discontinuous dielectric interfaces, automatic differentiation may lead to accuracy loss, high-frequency oscillations, and convergence difficulties \cite{Hu2025}, thereby hindering the efficient convergence of the residuals of the partial differential equation (PDE). Moreover, key questions, such as how to effectively handle electromagnetic physical boundaries and utilize boundary information to accelerate convergence, how to perform accurate simulations for electromagnetic modes with different polarizations, e.g., transverse-magnetic (TM) or transverse-electric (TE) modes, and whether a standardized computational framework can be developed to obtain stable and reliable computational results, are yet to be addressed.

In this paper, we consider two-dimensional (2D) electromagnetic scattering problems as a generic example and construct a unified PINN-based computational framework within which to address them. We apply this computational framework to electromagnetic modes with different polarizations and validate it across various scattering scenarios, ranging from single-dielectric to complex multi-dielectric scatterers. Then, the predictions of our numerical method are compared to analytical and/or numerical solutions obtained using well-tested methods such as the FDTD method. An improved integration scheme for physical boundaries is proposed, particularly for the TE polarization; the accuracy of the improved PINN-based method applied to complex electromagnetic scattering configurations is consistently achieved. Our article is organized as follows. In Sec~\ref{theory} we present the theoretical framework on which our method is based, then in Sec.~\ref{methods}, we illustrate how it can be implemented for the cases of the TM and TE polarizations. Furthermore, in Sec.~\ref{examples}, we demonstrate the versatility of our method by applying it to several scattering configurations of various degrees of complexity. Finally, in the last section, we outline the key conclusions of our study.



\section{Theoretical Foundations for PINNs}\label{theory}
The main idea behind PINN-based methods lies in using neural networks to approximately solve physical equations by minimizing a loss function that includes residuals of governing equations and boundary conditions, so as to ensure that the physics of the problem is consistently satisfied. The typical architecture of a PINN consists of an input layer, several hidden layers, and an output layer. The input layer receives the spatial coordinates and related physical parameters, and by embedding Maxwell equations and boundary conditions into the loss function, the output layer generates the target physical field. Importantly, it can effectively capture the distribution and propagation characteristics of electromagnetic fields without requiring explicit mesh generation.
\begin{figure*}[!t]
\centering
\includegraphics[width=\textwidth]{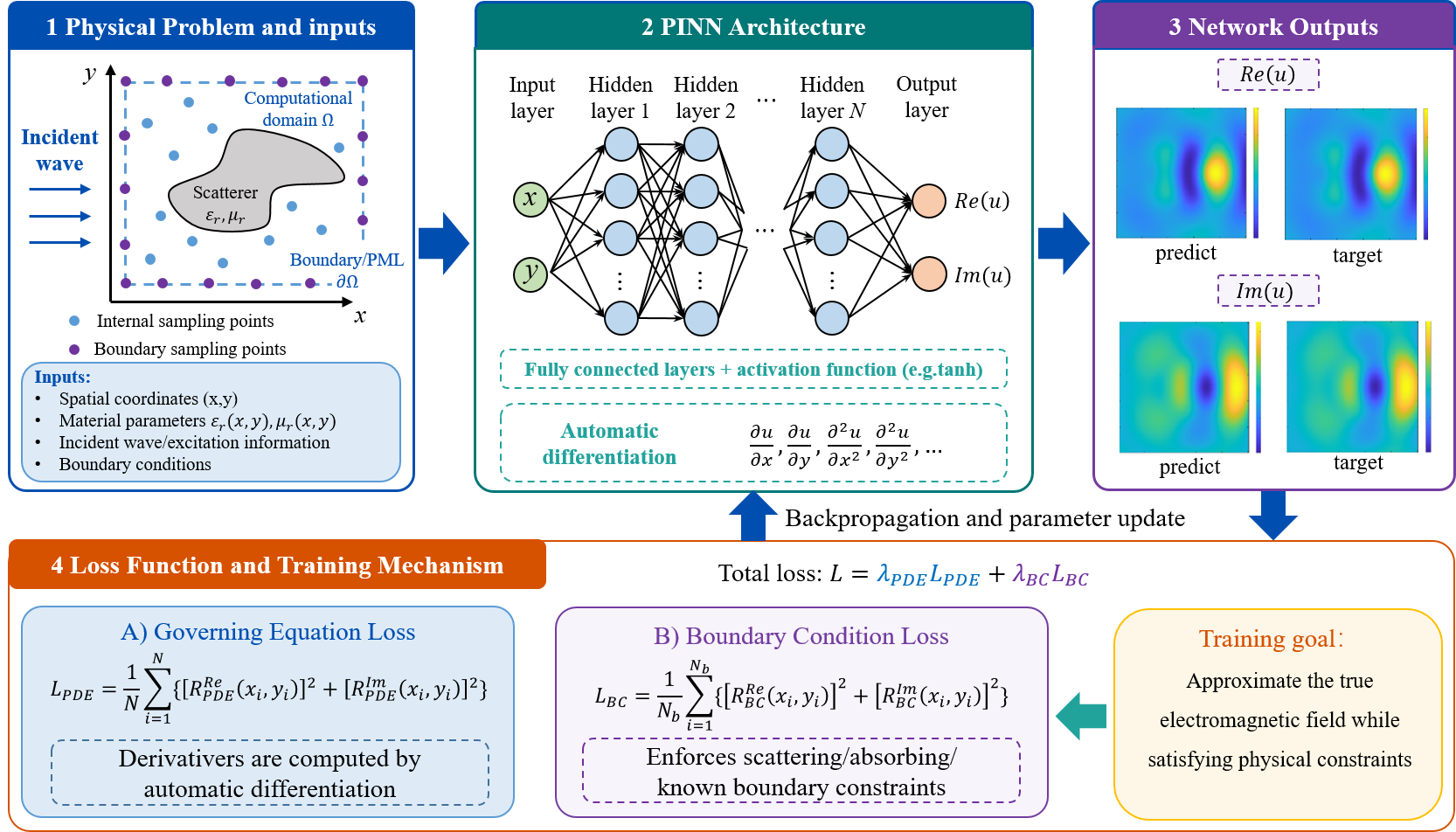}
\caption{Schematic illustration of the PINN-based method. The input layer, which receives the spatial coordinates $x$ and $y$, is followed by a certain number of hidden layers and the automatic differentiation block. The constraints are defined by the physics residual loss function and the boundary conditions residual loss function, whereas the total loss function is defined as a weighted sum of the physics residual and boundary residual loss functions.}
\label{fig:PINN diagram}
\end{figure*}

These ideas are illustrated in Figure~\ref{fig:PINN diagram}, where we show the schematic diagram of a generic PINN used in this study. In electromagnetic scattering modeling, the choice of activation function directly affects the physical consistency and numerical stability of the model. In this study, we adopt as an activation function a hyperbolic tangent function \cite{Gnanasambandam2022}, which is bounded and has a continuous second-order derivative. It not only satisfies the high-order derivative requirements of the Helmholtz equation, but it also conforms to the physical characteristics of the electromagnetic fields, thus effectively capturing phase continuity in wave propagation and properly handling interface discontinuities.

In the training of PINNs, the loss function plays a key role by embedding into the optimization objective both the PDEs that describe the evolution of the fields and the corresponding boundary conditions. Typically, the objective function has two parts: the physics residual computed via automatic differentiation based on the network outputs for the governing PDEs, and the boundary residual that ensures that the boundary conditions are satisfied. This dual-constraint approach ensures that the PINN simultaneously satisfies the governing PDEs and boundary conditions, thereby predicting accurate and physically consistent solutions.

To strictly constrain the PDEs, the physics residual loss function is defined as the sum of the squared residuals, i.e.:
\begin{equation} \label{eq:PDE_loss}
\mathcal{L}_{PDE} = \frac{1}{N} \sum_{i=1}^{N} \left\{ \left[ \mathcal{R}_{PDE}^{\text{Re}}(x_i,y_i) \right]^2 + \left[ \mathcal{R}_{PDE}^{\text{Im}}(x_i,y_i) \right]^2 \right\}
\end{equation}
where $N$ is the number of sampling points in the training dataset, $(x_i,y_i)$, $i=1, \ldots, N$, are internal interpolation points, and $\mathcal{R}_{PDE}^\text{Re}$ ($\mathcal{R}_{PDE}^\text{Im}$) is the real (imaginary) part of the residual, $\mathcal{R}_{PDE}$.

The loss function pertaining to the boundary conditions ensures the well-posedness of the solution by constraining the output behavior of the neural network at the boundary of the computational domain. The following mathematical expression defines it:
\begin{equation} \label{eq:BC_loss}
   \mathcal{L}_{BC} = \frac{1}{N_{b}} \sum_{i=1}^{N_b} \left\{ \left[ \mathcal{R}_{BC}^\text{Re}(x_i,y_i) \right]^2 + \left[ \mathcal{R}_{BC}^{\text{Im}}(x_i,y_i) \right]^2 \right\}
\end{equation}
where $N_{b}$ denotes the number of boundary sampling points in the training dataset, $(x_i,y_i)$, $i=1, \ldots, N_{b}$, are the sampled points on the boundary, and $\mathcal{R}_{BC}^\text{Re}$ ($\mathcal{R}_{BC}^\text{Im}$) is the real (imaginary) part of the residual, $\mathcal{R}_{BC}$. Note that the mathematical expressions of the residual functions $\mathcal{R}_{PDE}$ and $\mathcal{R}_{BC}$ will be provided in the next section.

The total loss function is defined as a weighted sum of the physics residual and boundary residual:
\begin{equation} \label{eq:total_loss}
 \mathcal{L}(\theta) = \lambda_{\text{PDE}} \mathcal{L}_{\text{PDE}} + \lambda_{\text{BC}} \mathcal{L}_{\text{BC}}
\end{equation}
where $\lambda_{\text{PDE}}$ and $\lambda_{\text{BC}}$ are weighting coefficients used to balance the contributions of the two loss terms. If experimental or synthetic data exist, an additional data fitting term, $\lambda_{\text{Data}}$, can be introduced. By minimizing the total loss function, PINNs can simultaneously satisfy the governing equations and boundary conditions, thereby achieving high-accuracy prediction of the scattered electromagnetic fields \cite{Cuomo2022}.

In PINNs, the optimizer iteratively updates the weights and biases of the neural network so as to minimize the loss function composed of physical constraints and data fitting, thus ensuring that the network output approximates the true solution while satisfying the governing equations and boundary conditions. Commonly used optimizers include the limited-memory Broyden-Fletcher-Goldfarb-Shanno (L-BFGS) algorithm and Adam. The L-BFGS optimizer exhibits superlinear convergence characteristics for quasi-convex functions, making it particularly suitable for high-accuracy optimization of small- to medium-scale problems. The Adam optimizer, as an adaptive learning rate algorithm, dynamically adjusts the learning rate of each parameter based on gradient statistics, effectively alleviating gradient vanishing/explosion problems and improving training stability. In this study, the optimizer computes gradients through backpropagation and updates network parameters, gradually guiding the solution towards an accurate representation that satisfies the physical laws governing the scattering problem. A dynamic weighting strategy is also employed to balance the contributions of different loss terms, ensuring robustness and convergence efficiency under complex physical constraints.

\section{Methods}\label{methods}
Consider a 2D region with a non-uniform relative electric permittivity distribution, $\varepsilon_r(x,y)$. The incident field is a plane wave propagating along the $x$-axis, whose time dependence is described by the exponential factor $e^{-i\omega t}$, with $\omega$ being the angular frequency. Within this set-up, one can simultaneously account for both TE and TM polarizations. For the TE polarization, the electric field lies within the $(x,y)$ propagation plane, meaning $E_x$ and $E_y$ are nonzero and $E_z=0$, whereas the magnetic field has only one nonzero component, $H_z$, the in-plane components being equal to zero, $H_x=H_y=0$. For the TM polarization, the magnetic field lies within the $(x,y)$-plane, i.e. $H_x$ and $H_y$ are nonzero and $H_z=0$, whereas the electric field is normal to the $(x,y)$-plane, meaning $E_z$ is nonzero and $E_x=E_y=0$.

If we consider the TM polarization, the scattering problem reduces to finding the electric field component $E_z$ of the scattered field, as $H_x$ and $H_y$ can be derived from $E_z$. Under these circumstances, the nonzero components of the electromagnetic field satisfy the following relations:
\begin{equation}
\begin{cases}
H_x = \frac{1}{i\omega \mu_{0}} \frac{\partial E_z}{\partial y} \\
H_y = -\frac{1}{i\omega \mu_{0}} \frac{\partial E_z}{\partial x} \\
\frac{\partial H_y}{\partial x} - \frac{\partial H_x}{\partial y} = -i\omega \varepsilon(x,y) E_z
\end{cases}
\end{equation}
where we assumed that the electromagnetic scatterers are made of non-magnetic materials, namely $\mu=\mu_{0}$. Eliminating the magnetic field components from the above equations yields the 2D Helmholtz equation for $E_z$:
\begin{equation}\label{eq:TM}
\nabla^2E_z + k_0^2 \varepsilon_r(x,y) E_z= 0,
\end{equation}
where $k_0=\omega\sqrt{\mu_0\varepsilon_0}$ is the free-space wave number.

The case of the TE polarization is treated in a similar way. Thus, in this case, one has to solve for $H_z$, as $E_x$ and $E_y$ can subsequently be derived from this component. In particular, in this case, the governing equations are:
\begin{equation}
\begin{cases}
E_x = -\frac{1}{i\omega \varepsilon} \frac{\partial H_z}{\partial y} \\
E_y = \frac{1}{i\omega \varepsilon} \frac{\partial H_z}{\partial x} \\
\frac{\partial E_y}{\partial x} - \frac{\partial E_x}{\partial y} = i\omega \mu_{0} H_z
\end{cases}
\end{equation}
Similarly, the governing equation for $H_z$ is obtained as:
\begin{equation}\label{eq:TE}
\nabla \cdot \left[ \frac{1}{\varepsilon_r(x, y)} \nabla H_z \right] + k_0^2  H_z = 0
\end{equation}

Comparing \eqref{eq:TM} and \eqref{eq:TE}, we observe a key difference between these two equations: since $\varepsilon_r(x, y)$ is not a continuous function, the coefficients in \eqref{eq:TM} are bounded but discontinuous at the dielectric boundaries; on the other hand, the derivatives associated with the divergence operator in \eqref{eq:TE} imply that the coefficients in this equation are singular at the dielectric boundaries. As will be shown in what follows, these differences greatly affect the convergence characteristics of the PINN-based method introduced in this work.

To simplify the notation, subsequent equations will describe in a unified way both the TM and TE cases by denoting the function we need to solve for, namely either $E_z(x,y)$ or $H_z(x,y)$, by $u(x,y)$. According to the general formalism of electromagnetic wave scattering, the total field in a non-uniform medium can be expressed as the linear superposition of the known incident field and the unknown scattered field:
\begin{equation}\label{eq:tot_fied}
u_{\text{tot}} = u_{\text{inc}} + u_{\text{sc}}
\end{equation}
where $u_{\text{tot}}$ denotes the total field, $u_{\text{inc}}$ is the incident field defined as a plane wave propagating along the $x$-axis, $u_{\text{inc}}(x,y) = e^{i k_m x}$, and $u_{\text{sc}}$ is the scattered field induced by the presence of the inhomogeneous medium. Here, $k_m=\omega\sqrt{\mu_{0}\varepsilon_{0}\varepsilon_{rm}}=k_{0}\sqrt{\varepsilon_{rm}}$ is the wave number in the background material, with $\varepsilon_{rm}$ being the relative electric permittivity of the background medium. For the sake of simplicity, in this study we assume that the background medium is free space, so that $\varepsilon_{rm}=1$ throughout.

In the TM case, the incident plane wave is a solution of the 2D Helmholtz equation \eqref{eq:TM}, so that it satisfies the equation:
\begin{equation}\label{eq:inc_field_governing_equation}
\nabla^2 u_{\text{inc}} + k_0^2 u_{\text{inc}} = 0
\end{equation}

When the domain contains a non-uniform relative permittivity, a scattered field $u_{\text{sc}}$ is induced. Substituting the total field \eqref{eq:tot_fied} into \eqref{eq:TM}, taking into account \eqref{eq:inc_field_governing_equation}, and rearranging the terms, leads to the following equation for the TM-polarized scattered field:
\begin{equation}\label{eq:TM_sc}
\left[\frac{1}{k_0^2}\nabla^2+\varepsilon_r(x, y)\right] u_{\text{sc}} = -\left[\varepsilon_r(x, y)-1\right] u_{\text{inc}}
\end{equation}
The left-hand side (l.h.s.) of this equation contains spatial differential terms of the scattering field and a material modulation term, whereas the right-hand side (r.h.s.) represents the equivalent source term induced by dielectric perturbations. This source term forms the key component for physics-based residual calculation in subsequent neural network training.

The equation governing the scattered field in the TE case is obtained in a similar way and can be written as:
\begin{multline}\label{eq:TE_sc}
\left\{\frac{1}{k_0^2}\nabla\cdot\left[\frac{1}{\varepsilon_r(x,y)}\nabla\right]+1\right\}u_{\text{sc}} = \\
- \left\{\frac{1}{k_0^2}\nabla\cdot\left[\frac{1}{\varepsilon_r(x,y)}\nabla\right]
    +1 \right\}u_{\text{inc}}
\end{multline}

Both the scattered field, $u_{\text{sc}}$, and the incident field, $u_{\text{inc}}$, are expressed as complex-valued functions, whose expressions can be written as:
\begin{align}
&u_{\text{sc}}(x,y)= u_{\text{sc}}^{\text{Re}}(x,y) + i u_{\text{sc}}^{\text{Im}}(x,y) \\
&u_{\text{inc}}(x,y)= u_{\text{inc}}^{\text{Re}}(x,y) + i u_{\text{inc}}^{\text{Im}}(x,y)
\end{align}
where $u_{\text{sc}}^{\text{Re}}$/$u_{\text{inc}}^{\text{Re}}$ ($u_{\text{sc}}^{\text{Im}}$/$u_{\text{inc}}^{\text{Im}}$) denotes the real (imaginary) part of the scattered/incident field.

Given that $k_m = k_0 \sqrt{\varepsilon_{rm}}=k_0$, the incident field can be expressed as:
\begin{equation}\label{eq:inc_pw}
u_{\text{inc}}(x,y) = e^{i k_0 x} = \cos(k_0 x) + i\sin(k_0 x)
\end{equation}
Inserting this expression into \eqref{eq:TM_sc}, we get the equations governing the real and imaginary parts of the scattered TM field:
\begin{equation}\label{eq:TM_PINN}
\left\{\begin{array}{l}
\left[\frac{1}{k_0^2}\nabla^2 + \varepsilon_r(x, y)\right] u_{\text{sc}}^{\text{Re}} = -\left[ \varepsilon_r(x, y) - 1 \right] \cos(k_0 x) \\
\left[\frac{1}{k_0^2}\nabla^2 + \varepsilon_r(x, y)\right] u_{\text{sc}}^{\text{Im}} = -\left[ \varepsilon_r(x, y) - 1 \right] \sin(k_0 x)
\end{array}
\right.
\end{equation}

Similarly, in the case of the TE polarization, the set of equations governing the real and imaginary parts of the scattered field can be obtained by substituting \eqref{eq:inc_pw} into \eqref{eq:TE_sc} and separating the real and imaginary parts:
\begin{equation}\label{eq:TE_PINN}
\left\{\begin{array}{l}
\left\{\frac{1}{k_0^2}\nabla\cdot\left[\frac{1}{\varepsilon_{r}(x,y)} \nabla \right]
+ 1\right\}u_{\mathrm{sc}}^{\mathrm{Re}} = \\
~~~~~\left(\frac{1}{\varepsilon_r}-1\right)\cos(k_0 x)
+ \frac{1}{k_0}\frac{\partial(1/\varepsilon_r)}{\partial x}\sin(k_0 x) \\
\left\{\frac{1}{k_0^2}\nabla\cdot\left[\frac{1}{\varepsilon_{r}(x,y)} \nabla \right]
+ 1\right\} u_{\mathrm{sc}}^{\mathrm{Im}} = \\
~~~~~\left(\frac{1}{\varepsilon_r}-1\right)\sin(k_0 x)
- \frac{1}{k_0}\frac{\partial(1/\varepsilon_r)}{\partial x}\cos(k_0 x)
\end{array}
\right.
\end{equation}

To incorporate \eqref{eq:TM_PINN} and \eqref{eq:TE_PINN} into the PINNs framework, corresponding residual functions must be constructed so as to define physics-based loss terms. These residual functions quantify the difference between the solution predicted by the network and that of the governing equations, thus serving as a key physical supervision quantity during the model training. For the TM polarization, the real and imaginary parts of the residual are defined, respectively, as follows:
\begin{equation}\label{eq:TM_Res_PDE}
\left\{\begin{array}{l}
\mathcal{R}_{PDE}^{\text{Re}}(x,y) = \left[\frac{1}{k_0^2}\nabla^2 + \varepsilon_r(x, y)\right] u_{\text{sc}}^{\text{Re}}  \\ ~~~~~+ \left[\varepsilon_r(x,y) - 1\right]  \cos(k_0x) \\
\mathcal{R}_{PDE}^{\text{Im}}(x,y) = \left[\frac{1}{k_0^2}\nabla^2 + \varepsilon_r(x, y)\right] u_{\text{sc}}^{\text{Im}}  \\ ~~~~~+ \left[\varepsilon_r(x,y) - 1\right] \sin(k_0x)
\end{array}
\right.
\end{equation}

Similarly, the real and imaginary parts of the residual, in the case of the TE polarization, can be expressed as:
\begin{equation}\label{eq:TE_Res_PDE}
\left\{\begin{array}{l}
\mathcal{R}_{PDE}^{\text{Re}}(x, y)=\left\{\frac{1}{k_0^2}\nabla\cdot\left[\frac{1}{\varepsilon_{r}(x,y)} \nabla \right]
+ 1\right\}u_{\mathrm{sc}}^{\mathrm{Re}} \\
~~~~~-\left(\frac{1}{\varepsilon_r}-1\right)\cos(k_0x)-\frac{1}{k_0}\frac{\partial(1/\varepsilon_r)}{\partial x}\sin(k_0x) \\
\mathcal{R}_{PDE}^{\text{Im}}(x, y)
= \left\{\frac{1}{k_0^2}\nabla\cdot\left[\frac{1}{\varepsilon_{r}(x,y)} \nabla \right]
+ 1\right\} u_{\mathrm{sc}}^{\mathrm{Im}} \\
~~~~~-\left(\frac{1}{\varepsilon_r}-1\right)\sin(k_0x)+\frac{1}{k_0}\frac{\partial(1/\varepsilon_r)}{\partial x}\cos(k_0x)
\end{array}
\right.
\end{equation}

Using \eqref{eq:TM_Res_PDE} and \eqref{eq:TE_Res_PDE} for the residuals corresponding to the TM and TE polarizations, respectively, one can construct the physical information loss function, $\mathcal{L}_{PDE}$, for the neural network. It is defined in \eqref{eq:PDE_loss} as the mean squared error (MSE) of the residuals computed at all internal training points, i.e., the average of the squared residuals computed at discrete interpolation points. During training, this loss function serves as the optimization objective, guiding the network outputs to progressively better satisfy the physical governing equations.

The physics of the scattering problem can be better understood if one uses the so-called equivalent sources. They are represented by the r.h.s. of \eqref{eq:TM_PINN} and \eqref{eq:TE_PINN}. For the TM polarization, the real and imaginary parts of the source term are:
\begin{equation}\label{eq:TM_source}
\left\{\begin{array}{l}
J^{\text{Re}}(x,y) = -\left[ \varepsilon_r(x,y) - 1 \right] \cos(k_0 x) \\
J^{\text{Im}}(x,y) = -\left[ \varepsilon_r(x,y) - 1 \right] \sin(k_0 x)
\end{array}
\right.
\end{equation}
whereas in the case of the TE polarization, they are expressed as:
\begin{equation}\label{eq:TE_source}
\left\{\begin{array}{l}
J^{\text{Re}}(x,y) = \left(\frac{1}{\varepsilon_r}-1\right)\cos(k_0 x)
+ \frac{1}{k_0}\frac{\partial(1/\varepsilon_r)}{\partial x}\sin(k_0 x) \\
J^{\text{Im}}(x,y) = \left(\frac{1}{\varepsilon_r}-1\right)\sin(k_0 x)
- \frac{1}{k_0}\frac{\partial(1/\varepsilon_r)}{\partial x}\cos(k_0 x)
\end{array}
\right.
\end{equation}

By embedding the source function into a physics-based residual, the PINNs framework explicitly captures the modulation effects induced by the non-uniformity of the media on electromagnetic fields. Consequently, the physical consistency of the network and its prediction accuracy are effectively enhanced when modeling wave scattering phenomena.

To ensure that one solves for outgoing radiating solutions of the scattering problem, appropriate radiation boundary conditions must be applied at the boundary of the computational domain. This amounts to the simulation of wave propagation in open space. For a time-harmonic problem with angular frequency $\omega$, the scattered field $u_{\text{sc}}$ must satisfy at infinity the radiation condition, also called the Sommerfeld boundary condition. The mathematical expression for this condition is:
\begin{equation}
\lim_{r \to \infty} r^{1/2} \left(\frac{\partial }{\partial r} - ik_0 \right)u_{\text{sc}} = 0
\end{equation}
where $r=\sqrt{x^{2}+y^{2}}$ denotes the distance from the origin. The Sommerfeld condition ensures that at infinity the scattered wave is an outgoing cylindrical wave, thereby preventing reflected waves from re-entering the computational domain. However, directly handling an infinite computational domain is not possible in practical numerical simulations, so a finite computational domain must be used \cite{Medvedev2025}\cite{Yu2024}. To balance computational efficiency and boundary absorption performance, in this study, we employ the following approximate local radiation condition at the computational domain boundary:
\begin{equation}
\frac{\partial u_{\text{sc}}}{\partial n} = ik_0u_{\text{sc}}
\end{equation}
where $\partial / \partial n$ denotes the derivative along the direction of the outward normal to the boundary. This condition assumes that the wave behavior near the boundary approximates a locally outward-propagating plane wave, constituting a first-order approximation of the radiation boundary condition.

Within the rectangular computational domain $\Omega = [x_{\min}, x_{\max}] \times [y_{\min}, y_{\max}]$, the absorption boundary condition can be expressed on the four boundaries as:
\begin{equation}
\left\{\begin{array}{l}
\left(\frac{\partial}{\partial x}+i k_0\right)u_{\text{sc}}\left.\right\vert_{x=x_{\min}} = 0 \\
\left(\frac{\partial}{\partial x}-i k_0\right)u_{\text{sc}}\left.\right\vert_{x=x_{\max}} = 0 \\
\left(\frac{\partial}{\partial y}+i k_0\right)u_{\text{sc}}\left.\right\vert_{y=y_{\min}} = 0 \\
\left(\frac{\partial}{\partial y}-i k_0\right)u_{\text{sc}}\left.\right\vert_{y=y_{\max}} = 0
\end{array}\right.
\end{equation}

To incorporate the above boundary conditions in the PINNs framework, they must be expressed as residual functions of the scattered field. Taking the right boundary, $x = x_{\max}$, as an example, whereby the outward normal points in the positive direction of the $x$-axis, the absorption boundary condition can be expressed as:
\begin{equation}
\frac{\partial}{\partial x} \left( u_{\text{sc}}^{\text{Re}} + i u_{\text{sc}}^{\text{Im}} \right)- ik_{0} \left( u_{\text{sc}}^{\text{Re}} + i u_{\text{sc}}^{\text{Im}} \right) = 0
\end{equation}
Separating the real and imaginary parts in this relation yields:
\begin{equation}
\begin{cases}
\displaystyle \frac{\partial u_{\text{sc}}^{\text{Re}}}{\partial x} + k_{0} u_{\text{sc}}^{\text{Im}} = 0 \\
\displaystyle \frac{\partial u_{\text{sc}}^{\text{Im}}}{\partial x} - k_{0} u_{\text{sc}}^{\text{Re}} = 0
\end{cases}
\end{equation}

To express in a unified way the residuals on all four boundaries, we introduce the sign variable, $\sigma \in \{-1, +1\}$, which denotes the sign of the direction of the normal derivative. The value of $\sigma$ depends on the boundary orientation, namely, $\sigma = -1$ at $x = x_{\min}$ and $y = y_{\min}$, whereas $\sigma = +1$ if $x = x_{\max}$ and $y = y_{\max}$. Using this notation, the real and imaginary parts of the residuals of the absorbing boundary condition can be expressed as:
\begin{equation}\label{eq:Res_BC}
\left\{\begin{array}{l}
\mathcal{R}_{\text{BC}}^{\text{Re}} = \frac{1}{k_{0}}\frac{\partial u_{\text{sc}}^{\text{Re}}}{\partial n} + \sigma u_{\text{sc}}^{\text{Im}} \\
\mathcal{R}_{\text{BC}}^{\text{Im}} = \frac{1}{k_{0}}\frac{\partial u_{\text{sc}}^{\text{Im}}}{\partial n} - \sigma  u_{\text{sc}}^{\text{Re}}
\end{array}
\right.
\end{equation}
Note that both these residual functions, as well as those in \eqref{eq:TM_Res_PDE} and \eqref{eq:TE_Res_PDE}, are dimensionless quantities. This facilitates the comparison of the efficiency and accuracy of the PINN method across different cases.

Within the PINNs framework, boundary conditions are incorporated into the training objective via the squared residual loss function $\mathcal{L}_{BC}$ defined in \eqref{eq:BC_loss}, thereby constraining the prediction accuracy of the network at domain boundaries. During training, the boundary loss function is optimized together with the residual loss function defined by the governing equations so as to ensure that the predictions of the network satisfy both the physical laws governing the scattering process inside the computational domain and the corresponding boundary radiation conditions.

\section{Numerical Experiments and Analysis}\label{examples}
To validate and investigate the accuracy and effectiveness of the proposed PINN method in solving 2D electromagnetic scattering problems for both TM and TE polarizations, a series of numerical experiments was designed, namely a single cylinder, a set of concentric cylinders, arbitrarily distributed cylinders, and irregular scatterers. The performance of our PINN method proposed in this paper was comprehensively evaluated by focusing on three aspects: accuracy, physical consistency, and geometric adaptability. All simulations were conducted under a unified computational framework, using consistent settings and training strategies. The performance of this method applied to different structural configurations was assessed by comparing the predictions of the PINN method with analytical solutions or with those obtained using the FDTD method.

In all cases investigated here, the computational domain was defined as a 2D rectangular region, $\Omega = [-1, 1] \times [-1, 1]$, with absorbing boundary conditions applied at all boundaries. The background medium was air, and we defined the relative permittivity distribution function, $\varepsilon_r(x,y)$, inside the computational domain as a piecewise continuous function that is different from unity only inside the scatterers. The incident wave was a time-harmonic plane wave propagating along the $x$-axis, with free-space wavelength $\lambda_{0} = 1$, corresponding to a wave number $k_0 = 2\pi/\lambda_{0} = 2\pi$. Note that since this work addresses mostly applications pertaining to the optical frequencies domain, we assumed that the characteristic spatial length of the scattering problem is of the scale of a micrometer, so that the unit length is $1\,\mu\text{m}$.
\begin{figure}[t]
\centering
\includegraphics[width=\columnwidth]{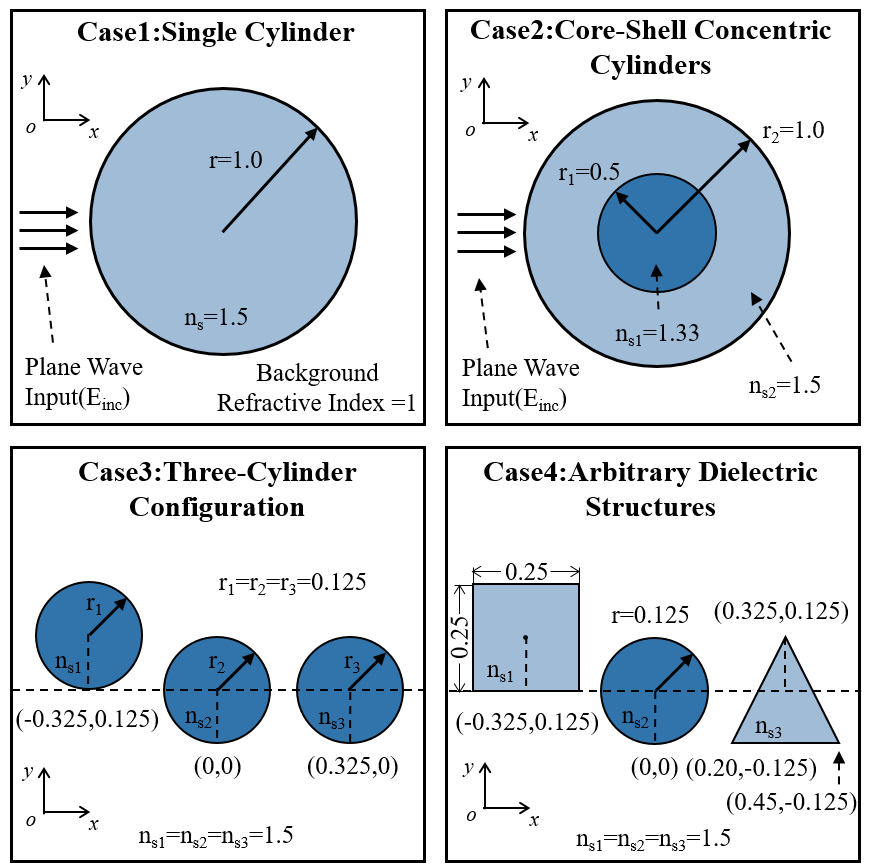}
\caption{Configurations of dielectric scatterers. \textit{Case 1}: single cylinder; \textit{Case 2}: concentric cylinders; \textit{Case 3}: three cylinders with different dielectric properties; and \textit{Case 4}: distribution of scatterers with different shape.}
\label{fig:example}
\end{figure}

The neural network architecture consists of 8 fully connected layers, each containing 20 neurons, with hyperbolic tangent activation functions. Network weight initialization employs the Glorot uniform distribution. Training samples were generated via Latin hypercube sampling (LHS), comprising 3000 interior collocation points, 100 boundary points (25 points per edge), and 50 auxiliary initial points. Training was performed using the Adam optimizer with a learning rate of $10^{-3}$. The total loss function is the sum of partial differential equation residuals and boundary residuals, with a dynamic weighting mechanism employed to enhance training stability. To ensure convergence and numerical stability of the training process, the maximum number of training iterations was set to $5 \times 10^{4}$. Additionally, the following dual early stopping criterion was adopted: training terminated when both the gradient norm and step size norm would fall below $10^{-5}$. This configuration effectively prevents overfitting and improves the convergence efficiency and robustness of the model in solving scattering problems.

In Figure~\ref{fig:example}, we schematically illustrate the configurations used in our numerical experiments. Thus, we considered four cases: (1) a single cylinder; (2) a core-shell system of concentric cylinders; (3) a set of arbitrarily located cylinders; and (4) a set of arbitrary scatterers. The first three cases are regular structures that can be validated by comparison with analytical solutions, whereas in the fourth case, no analytical solution exists, so the validation of our PINN-based approach was performed by using the FDTD method.
\begin{figure*}[!ht]
\centering
\includegraphics[width=\textwidth]{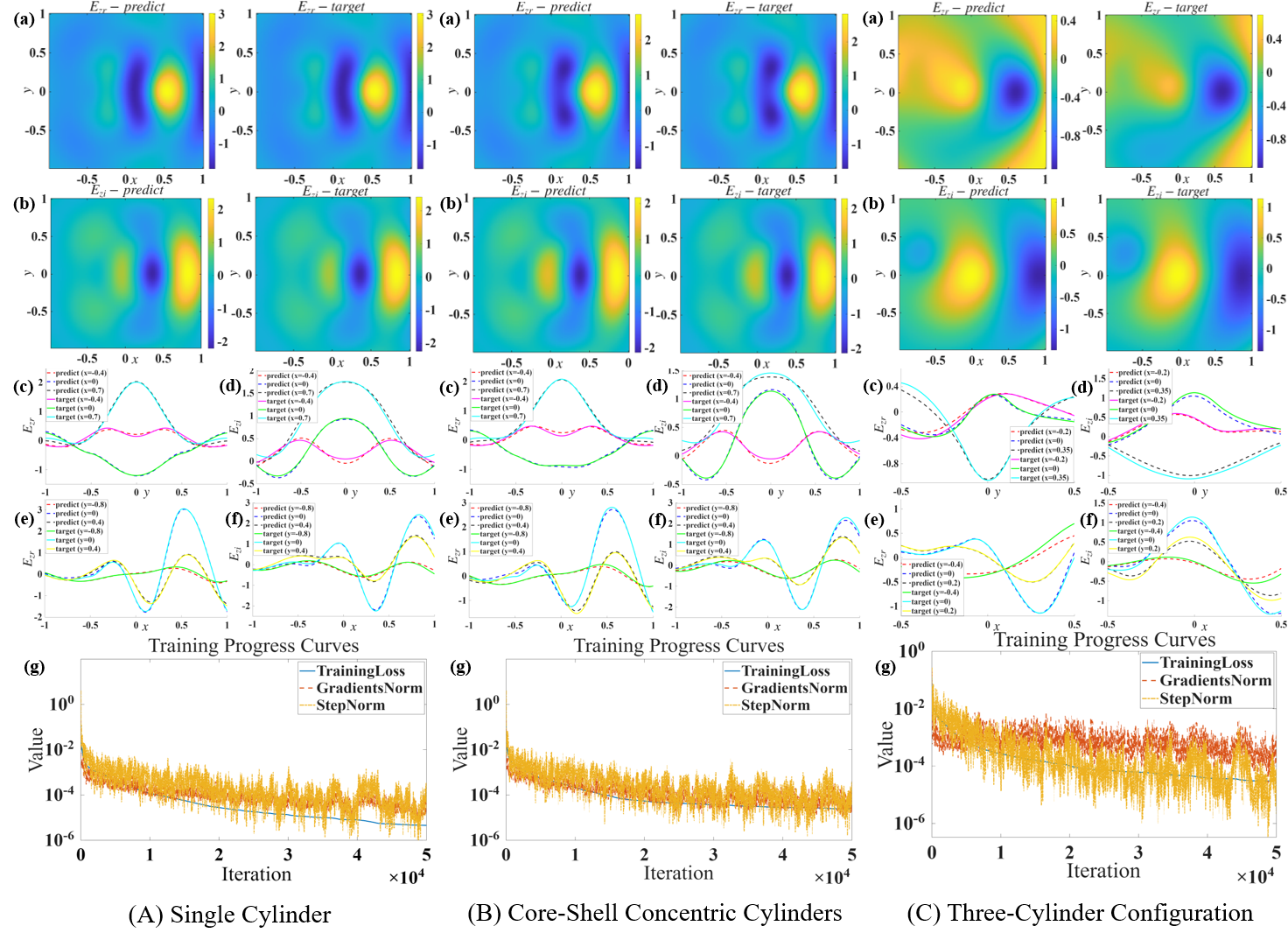}
\caption{Comparison of PINN predictions with analytical solutions under TM polarization.
(A) single cylinder; (B) core-shell concentric cylinders; (C) three-cylinder configuration.
Subfigure descriptions:
(a) real part $E_{zr}$ for $E_z$;
(b) imaginary part $E_{zi}$ for $E_z$;
(c), (d) distributions of $E_{zr}$ and $E_{zi}$, respectively, along three vertical lines at $x=-0.4$, $0$, and $0.7$;
(e), (f) distributions of $E_{zr}$ and $E_{zi}$, respectively, along four horizontal lines at $y=-0.8$, $0$, and $0.4$;
(g) loss function curve of PINNs.}
\label{fig:TM}
\end{figure*}

\begin{figure*}[!ht]
\centering
\includegraphics[width=\textwidth]{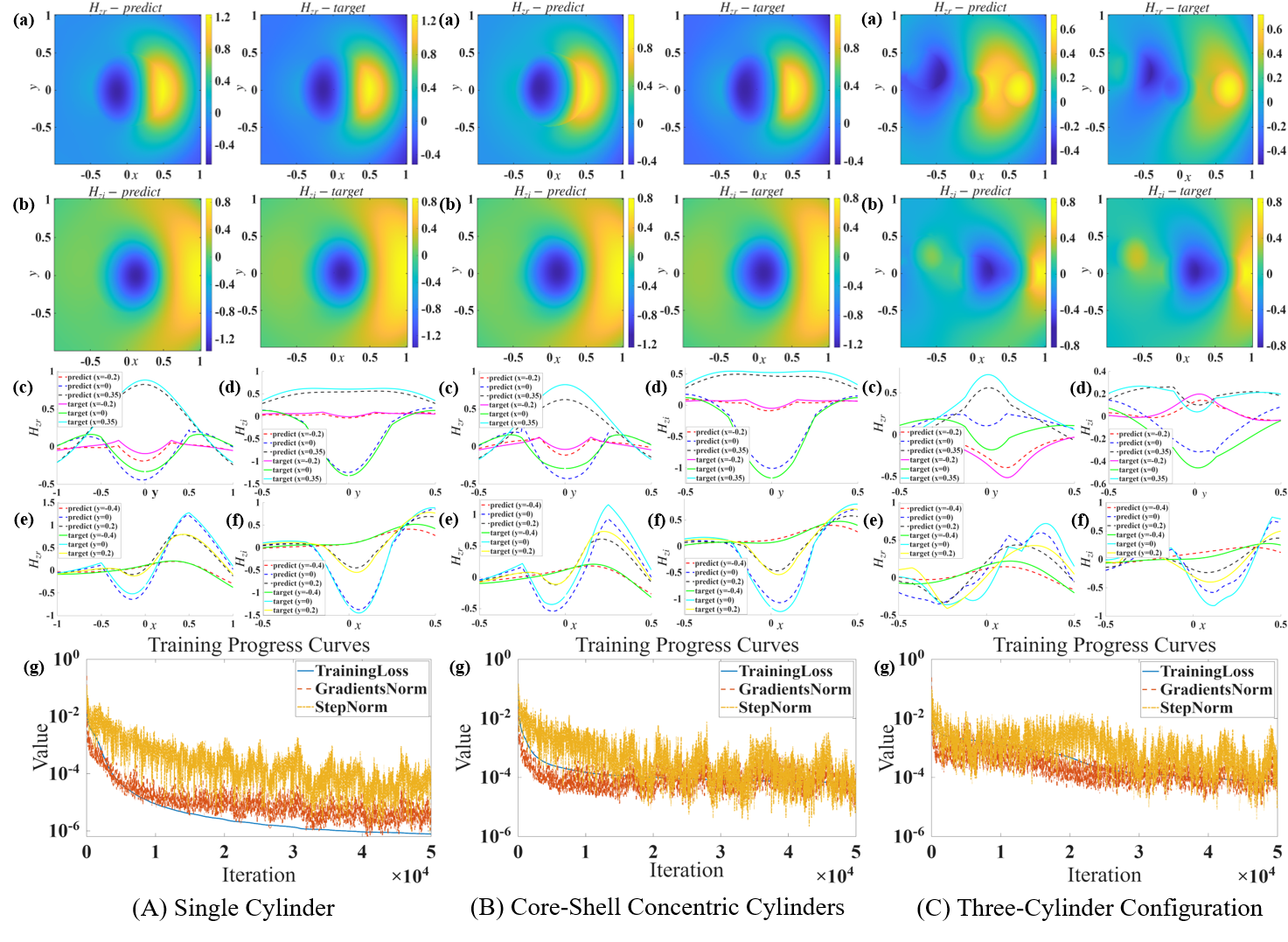}
\caption{Comparison of PINN predictions with analytical solutions under TE polarization.
(A) single cylinder; (B) core-shell concentric cylinders; (C) three-cylinder configuration.
Subfigure descriptions:
(a) real part $E_{zr}$ for $E_z$;
(b) imaginary part $E_{zi}$ for $E_z$;
(c), (d) distributions of $E_{zr}$ and $E_{zi}$, respectively, along three vertical lines at $x=-0.2$, $0$, and $0.35$;
(e), (f) distributions of $E_{zr}$ and $E_{zi}$, respectively, along four horizontal lines at $y=-0.4$, $0$, and $0.2$;
(g) loss function curve of PINNs.}
\label{fig:TE}
\end{figure*}

\subsection{Dielectric Structures Containing Cylinders or Core-shell Layered Cylinders}
In this subsection, we systematically investigate the accuracy and versatility of the PINN-based method applied to three benchmark cases of increasing complexity. First, we consider a dielectric cylinder with radius $r=1$ and refractive index $n_s=1.5$, located at the origin of the coordinate system. The spatial distribution of the relative permittivity is:
\begin{equation}\label{eq:one_cyl}
    \varepsilon_r(x,y) =
    \begin{cases}
    n_s^2, & \sqrt{x^2 + y^2} \leq r \\
    1,  & \sqrt{x^2 + y^2} > r
    \end{cases}
\end{equation}

To investigate the effects introduced by the inhomogeneity of the medium, we also considered a core-shell structure comprised of an inner cylinder with radius, $r_{1}=0.5$, and a concentric cylindrical shell with outer radius, $r_{2}=1$. The spatial distribution of relative permittivity is:
\begin{equation}\label{eq:core_shell}
\varepsilon_r(x,y) =
\begin{cases}
n_{s1}^2, & \sqrt{x^2 + y^2} \leq r_1 \\
n_{s2}^2, & r_1 < \sqrt{x^2 + y^2} \leq r_2 \\
1,  & \sqrt{x^2 + y^2} > r_2
\end{cases}
\end{equation}
where $n_{s1}$ and $n_{s2}$ denote the refractive indices of the core and shell regions, respectively, and are specified as $n_{s1}=1.33$ and $n_{s2}=1.5$ in this study.

To further analyze the versatility of the proposed PINN-based method, we considered a multiple-scatterers problem, namely a configuration consisting of 3 cylinders. The cylinders have the same radius and do not overlap. The coordinates, $(x_{ci},y_{ci})$, $i=1,2,3$, of the center of the left, middle, and right cylinders are $(-1.3d,0.5d)$, $(0,0)$, and $(1.3d,0)$, respectively, where $d = 0.25$ represents the diameter of the cylinders. The cylinders are assumed to have the same index of refraction, that is $n_{s1} = n_{s2} = n_{s3} = 1.5$, whereas the remaining simulation parameters are the same as in the previous two cases. Under these circumstances, the spatial distribution of the relative permittivity is:
\begin{equation}\label{eq:three_cyl}
\varepsilon_r(x,y) =
\begin{cases}
n_{si}^2, & (x - x_{ci})^2 + (y - y_{ci})^2 \leq r^2, i=1,2,3 \\
1,    & \text{otherwise}
\end{cases}
\end{equation}
where $r=d/2=0.125$ represents the radius of the cylinders.

In the following, we discuss the simulation results obtained in these three representative cases. Thus, in Figures~\ref{fig:TM} and \ref{fig:TE} we present a comparison between the predictions of our PINN-based method and the corresponding results obtained analytically for the TM and TE polarizations, respectively. In both figures, (A), (B), and (C) correspond to the case of a single cylinder, the case of concentric core-shell cylindrical layers, and the case of 3 cylinders, respectively.

Panels (a) and (b) of Figure~\ref{fig:TM} show the spatial distributions of the real and imaginary parts of the electric field component, $E_z$, respectively, with both the PINN-predicted results (left) and analytical solutions (right) being depicted. For a more facile comparison, the color maps have the same scale, with color bars indicating the value of the field magnitudes within the computational domain. It can be inferred from these plots that the PINN-predicted spatial distributions of both the real and imaginary parts of $E_z$ are in excellent agreement with the analytical solutions. Under different configurations of the scatterers, the PINNs model effectively captures the main physical characteristics of the field distributions, such as local field enhancement and field variations near the boundaries of the scatterers.

To further illustrate the accuracy of our method, we plot the real and imaginary parts of the electric field, calculated along the vertical lines defined by $x=-0.4$, $0$, $0.7$ and horizontal lines defined by $y=0.1$, $0.5$, $0.7$, $0.9$, and then compare the results with the analytical solutions. Panels (c) and (d) in Figure~\ref{fig:TM} show the distributions of the real and imaginary parts of $E_z$ along the three vertical cutlines, whereas panels (e) and (f) in the same figure show the distributions along the four horizontal cutlines. The results indicate that the predicted field distributions agree well at most positions with the analytical solutions. The best agreement is achieved for the single cylinder and core-shell structures. When abrupt changes in the index of refraction occur and when there are electromagnetic interactions among the scatterers, slight deviations appear in the regions of large gradients of $\varepsilon_{r}(x,y)$. However, overall our method performs well, demonstrating that PINNs can effectively predict the scattering field.

An important characteristic of a PINN-based numerical method is its convergence properties. Thus, to analyze the convergence performance of our numerical method during the training phase, we determined evolution curves of the training loss and the norms of the gradient and step size. The dependence of these quantities on the iteration number is plotted in the panels (g) of Figure~\ref{fig:TM}. These plots clearly demonstrate the dynamic characteristics of the model convergence process. As shown in these panels, the training loss rapidly decreases in the early stage of the training, then gradually stabilizes, and finally converges to a low value, which indicates that the network has effectively learned the physical constraints embedded in the governing equations and boundary conditions. Both the norms of the gradient and the step size exhibit steady declining trends without large oscillations or divergence phenomena, reflecting good numerical stability of the optimization process, parameter update regulation, and overall stability. The dynamic weighting mechanism introduced for the residual terms significantly enhances the efficiency of multi-objective optimization. No gradient explosion or overfitting phenomena occurred throughout the training process, proving that the model possesses good trainability and numerical robustness. A similar analysis of the results corresponding to the TE polarization, summarized in Figure~\ref{fig:TE}, reveals that the loss functions in all test cases stably converge within a finite number of iterations, achieving an accuracy of the order of $10^{-4}$.

The relative $L^{2}$ error is employed to quantitatively evaluate the discrepancy between the PINN-predicted solution and the reference solution, which is defined as
\begin{equation}
\varepsilon_{L_2}=\frac{\sqrt{\sum_{i=1}^{N}\left|E_{z,i}^{\mathrm{pred}}-E_{z,i}^{\mathrm{ref}}\right|^2}}{\sqrt{\sum_{i=1}^{N}\left|
E_{z,i}^{\mathrm{ref}}\right|^2}}
\label{eq:l2_error}
\end{equation}
Here, $E_{z,i}^{\mathrm{pred}}$ and $E_{z,i}^{\mathrm{ref}}$ denote the predicted and reference values of the complex electric field at the $i$th testing point, respectively, and $N$ is the total number of testing points.

\begin{table}[t!]
\centering
\caption{Relative $L^{2}$ Errors}\label{t:comp}
\begin{tabular}{>{\centering\arraybackslash}p{4cm}cc}
\toprule
\textbf{Scatterer type} & \textbf{TM polarization} & \textbf{TE polarization} \\
\midrule
single cylinder & 0.0708 & 0.9184 \\
core-shell concentric cylinders & 0.0804 & 0.1983 \\
three-cylinder configuration & 0.151 & 1.0936 \\
\bottomrule
\end{tabular}
\end{table}

A simple comparison of the $\varepsilon_{L_2}$ corresponding to the TM and TE polarizations, summarized by the data presented in Table~\ref{t:comp}, reveals that for the TM polarization, one achieves a better numerical accuracy across all scattering configurations. In the case of the TM polarization, the error increases only marginally with the structural complexity of the system of scatterers, underlying the robust modeling capability and versatility of our method. By contrast, in the case of the TE polarization, the errors are larger and depend much more strongly on the structural complexity of the scatterers.

The larger errors in the case of the TE polarization are primarily attributed to the mathematical form of the corresponding governing equations. More specifically, for the TE polarization, the governing equation \eqref{eq:TE} is particularly sensitive to abrupt variations of material properties at interfaces, as it contains partial derivatives of the discontinuous function ${1}/{\varepsilon_r(x,y)}$. As a result, mathematical singularities are induced at interfaces, whereas neural networks inherently seek solutions represented by smooth functions. By contrast, the equations governing the TM-polarized fields do not contain such singularities, so that in this case interfaces can be treated in a relatively simple manner, leading to an increased accuracy of our PINN-based numerical method.
\begin{figure}[t!]
    \centering
    \includegraphics[width=\columnwidth]{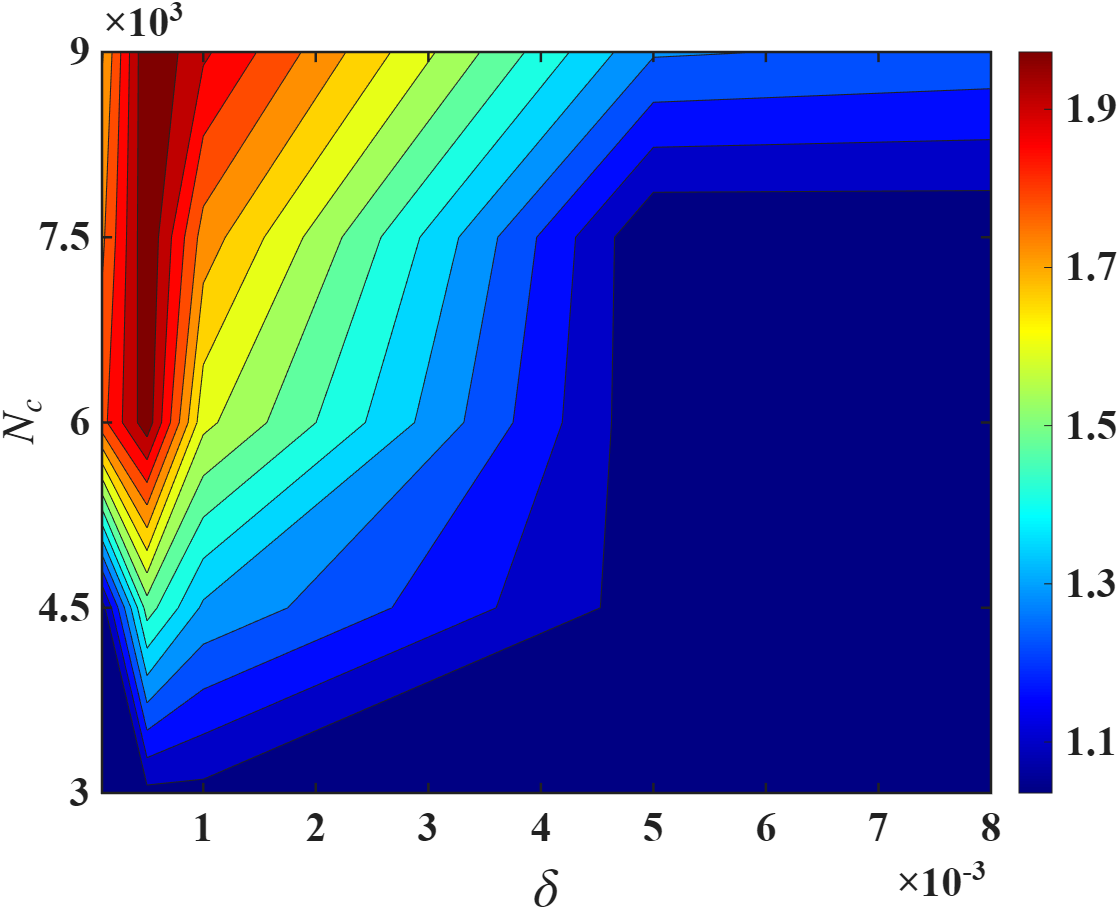}
    \caption{Dependence of $L^{2}$ error on the width of the transition region and number of sampling points.}
    \label{fig:err}
\end{figure}

To address this drawback, we introduce a smoothing function that removes the singularities at the boundaries. The specific implementation method is as follows: a transition region is defined near the interfaces of the scatterers, and the originally discontinuous permittivity distribution is smoothed out via the smoothing function. More specifically, let us assume that $P_{0}=(x_{0},y_{0})$ is a point located at the interface between the scatterer and background, and choose the width of the transition region to be equal to $\delta$. Then, if $\varepsilon_1$ ($\varepsilon_2$) is the electric permittivity inside (outside) the scatterer, the smoothed out permittivity is expressed as:
\begin{equation}\label{eq:smooth_eps}
\varepsilon_r(s) = \varepsilon_1 + \frac{1}{2}({\varepsilon_2 - \varepsilon_1})\left[1 + \tanh\left(\frac{s}{\delta}\right)\right]
\end{equation}
Here, $s$ is the distance to $P_{0}$ measured along the normal to the interface at $P_{0}$, defined in such a way that $s\geq0$ ($s<0$) outside (inside) the scatterer. This function ensures a smooth continuous variation of the permittivity across the transition region, thus removing the singularities associated with the derivatives of ${1}/{\varepsilon_r(x,y)}$.

To evaluate the effectiveness of this approach in improving the proposed method, we analyzed the dependence of the $\varepsilon_{L_2}$ on two key parameters: the number of sampling points $N_c$ and the characteristic width $\delta$ of the transition region. The analysis was performed for the most challenging configuration, i.e., the three-cylinder system. The corresponding results are shown in Fig.~\ref{fig:err}.

The first conclusion revealed by this figure is that the transition width has a more significant impact on the error. When the transition width is small, namely $\delta \le 10^{-3}$, the $\varepsilon_{L_2}$ is as large as 1.9. In this case, the model is overly sensitive to the large gradients present near the boundaries and fails to effectively capture the physics of the problem. In particular, when the transition region width is $\delta \approx 5 \times 10^{-4}$, the model breaks down, so that obtaining accurate results is not possible. When the transition width increases to about $5 \times 10^{-3}$, the error decreases significantly and enters a plateau region of values of about 1.1. Beyond $\delta \approx 5 \times 10^{-3}$, only marginal improvement is achieved by further widening the transition region, which indicates that the enhancement of the performance of the model by tuning the transition width has reached saturation.

Regarding the influence of the number of internal sampling points on the $\varepsilon_{L_2}$, Figure~\ref{fig:err} suggests that there are two parameter regions where the model exhibits markedly different properties. Thus, when the transition region is about $5 \times 10^{-3}$ or smaller, increasing the number of sampling points actually leads to increased error, which suggests that an excessive number of sampling points within a narrow transition region may introduce optimization bias or boost residuals in boundary regions with large curvature. On the other hand, when the transition region is sufficiently wide, namely $\delta \ge 5 \times 10^{-3}$, the impact of increasing the number of sampling points on the error gradually diminishes, indicating that the model has reached a stable state whereby continuing to increase the number of sampling points no longer improves its performance.

This analysis suggests that, for the TE polarization, special attention should be given in practical applications to selecting a suitable transition region width, thus balancing the stability of the model with the accuracy of the predicted results. As a final comment on this matter, we note that, since in this study the lengths are measured in $\mu\text{m}$, a characteristic width of $10^{-3}$ is in the range of nanometers, which is in fact the characteristic length over which the electric permittivity varies at the boundaries of dielectric media.

\subsection{Arbitrary Dielectric Structures}
In this subsection, we consider a more challenging configuration of scatterers, namely one for which analytical solutions do not exist. In particular, as per Figure~\ref{fig:example}, \textit{Case 4}, we consider a system containing a square, a cylinder, and a triangle. In this case, the benchmark solution is computed using the FDTD method. This configuration is characterized by a more intricate spatial distribution of the relative permittivity $\varepsilon_r(x, y)$, as in this case, not only that $\varepsilon_r(x, y)$ is not continuous at the boundaries of the scatterers, but the boundaries themselves are not smooth curves.

The specific geometric configuration of the scatterers is defined as follows. The cylinder is centered at $(0,0)$, has a diameter of 0.25, and has a refractive index of $n_{s3}$. The rectangular scatterer is centered at $(x_{c1}, y_{c1})=(-0.325,0.125)$, occupies the region $x_r\in[-0.45,-0.20]$ and $y_r\in[0,0.25]$, and has a refractive index of $n_{s1}$. The triangular scatterer is an isosceles triangle with vertices located at $(x_{t1}, y_{t1})=(0.325,0.125)$, $(x_{t2}, y_{t2})=(0.20,-0.125)$, and $(x_{t3}, y_{t3})=(0.45,-0.125)$, and has a refractive index of $n_{s2}$. The refractive indices of the rectangular, triangular, and cylindrical scatterers are chosen to be the same, that is $n_{s1}=n_{s2}=n_{s3}=1.5$.
\begin{figure}[t!]
	\centering
    \includegraphics[width=\columnwidth]{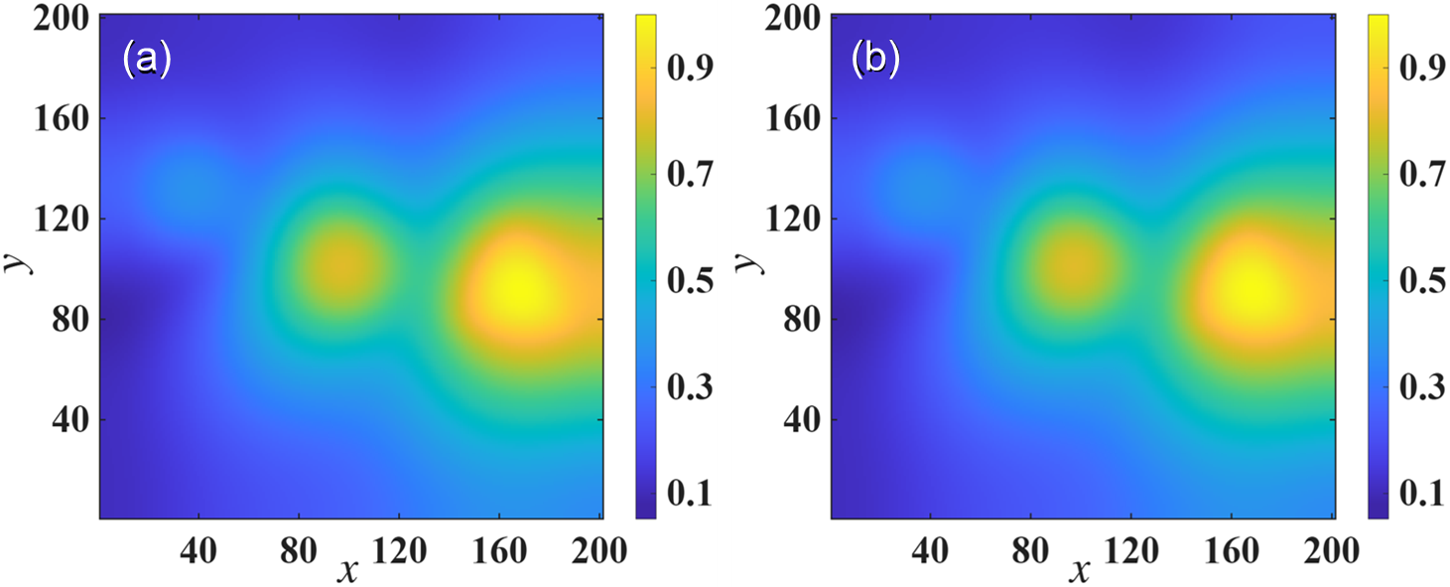}
	\caption{Spatial distribution of $E_{z}(x,y)$ corresponding to the TM polarization, computed using (a) the PINN-based method and (b) the FDTD-based method.}
	\label{fig:irr_TM}
\end{figure}

Figure~\ref{fig:irr_TM} shows the 2D distribution of the intensity of the electric field predicted by the PINN-based method and the benchmark solution calculated using the FDTD method. It can be seen from this figure that the PINN-based method accurately reproduces the key features of the scattered field distribution, and despite some minor amplitude deviations near boundaries, overall the agreement between the two solutions is very good. This indicates that for the TM polarization the proposed model performs rather well.

\begin{figure}[t!]
	\centering
	\includegraphics[width=\columnwidth]{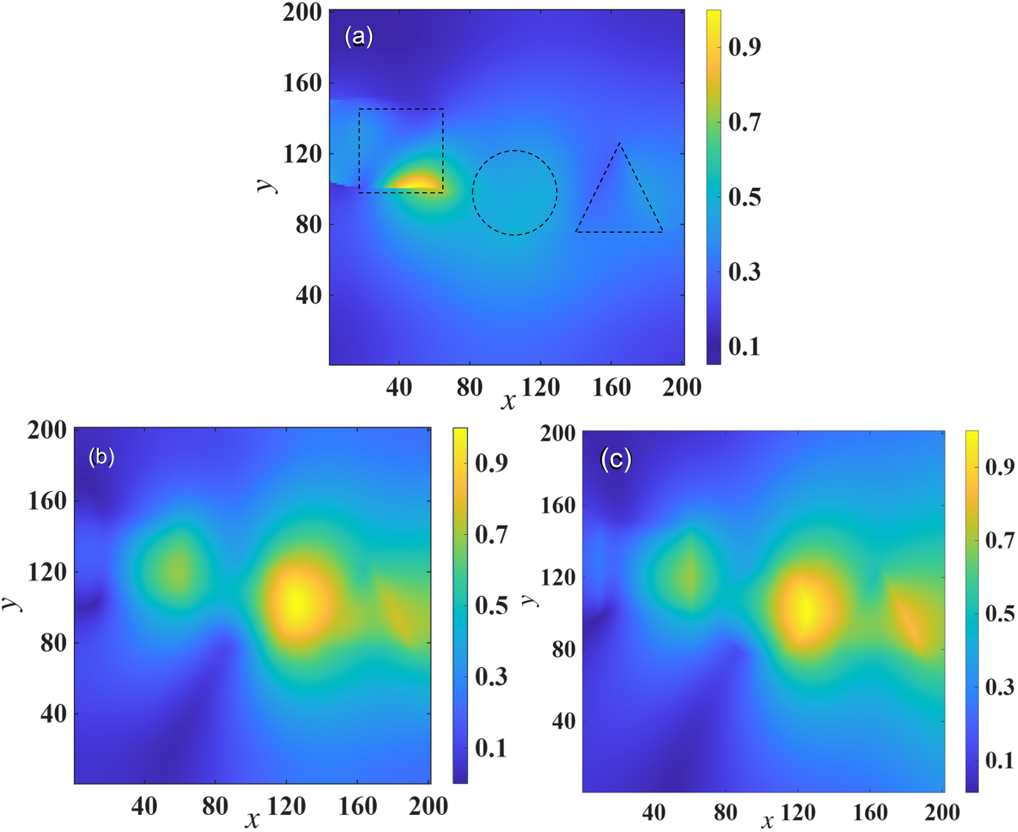}
	\caption{Spatial distributions of $H_{z}(x,y)$ corresponding to the TE polarization, computed using (a) the PINN-based method without smoothing function, (b) the PINN-based method with smoothing function, and (c) the FDTD-based method.}
	\label{fig:irr_TE}
\end{figure}

By contrast, in the case of the TE polarization, the PINN-based method produces reliable results only when a smoothing function is employed. This conclusion is supported by the results presented in Figure~\ref{fig:irr_TE}. Thus, the panels (A) and (B) of Figure~\ref{fig:irr_TE} show the scattered field distributions predicted by the PINN method before and after adding the smoothing function, respectively, whereas in panel (C) the benchmark solution obtained by using the FDTD method is plotted. In these computations, we used a combination of 3000 sampling points and a width of the smoothing region of $\delta=10^{-3}$, parameter values that minimize the value of $\varepsilon_{L_2}$.

If no smoothing function is used, the predictions of the PINN-based method are obviously incorrect, especially at the edges of the scatterers. On the other hand, after adding the smoothing function, the performance of the PINN method improves significantly, with the numerical errors near boundaries being noticeably reduced. This indicates that the introduction of the smoothing function enables the PINN to more accurately capture boundary conditions in systems with more complex scatterer configurations. However, compared to the benchmark solution, the results obtained using the PINN method still show minor deviations in certain regions, especially near boundaries. These differences may stem from the approximation capability of the network model and limitations of the method used to treat the boundaries. Overall though, the introduction of the smoothing function clearly leads to an improvement of the predictions of our PINN-based method to an extent at which similar levels of accuracy are achieved for both the TM and TE polarizations.

\section{Conclusion}
A physics-informed neural network framework is introduced for solving two-dimensional electromagnetic wave scattering problems in inhomogeneous dielectric media. The frequency-domain Maxwell equations and radiation boundary conditions are incorporated directly into the loss function, thus enabling the prediction of the scattered fields for both transverse magnetic and transverse electric polarizations within a meshless computational domain. Numerical experiments on several configurations of electromagnetic scatterers with increasing structural complexity demonstrate that the proposed approach accurately predicts field distributions with intricate spatial distributions.

For the case of transverse magnetic polarization, consistently small relative $L^{2}$ errors are observed, thus validating the accuracy of the proposed formulation. On the other hand, in the case of the transverse electric polarization, the discontinuity of $1/\varepsilon_r$ induces singularities in the equations governing the dynamics of the scattered fields near material interfaces, potentially degrading the validity of the predicted solutions. To address this issue, a smoothing technique employing a hyperbolic tangent is applied at the boundary regions, mitigating singular behavior and improving the agreement between the predictions of our method and the benchmark results obtained via the finite-difference time-domain method. Furthermore, the convergence trends of the loss function, gradient norm, and step-size norm confirm stable and efficient training across diverse configurations of the scatterers. These findings underscore the robustness and scalability of the proposed framework for electromagnetic scattering problems involving complex geometries and discontinuous material distributions. Future research will address three-dimensional geometries, advanced material models, and hybrid strategies combining PINNs with traditional numerical solvers.

\ifCLASSOPTIONcaptionsoff
  \newpage
\fi

\end{document}